\documentclass{jfm}
\usepackage{graphicx}
\usepackage{epstopdf, epsfig}

\usepackage[colorlinks, allcolors=blue]{hyperref} 

\newcommand\Ra{\mbox{\textit{Ra}}}
\newcommand\Cn{\mbox{\textit{Cn}}}

\newcommand\We{\mbox{\textit{We}}}

\newcommand\Nun{\mbox{\textit{Nu}}}
\newcommand\Prn{\mbox{\textit{Pr}}} 
\newcommand\Pe{\mbox{\textit{Pe}}}

\shorttitle{Enhancing Heat Transport in Multiphase RB Turbulence}
\shortauthor{H.-R. Liu, K. L. Chong, C. S. Ng, R. Verzicco and D. Lohse}

\title{Enhancing Heat Transport in Multiphase Rayleigh-B\'enard Turbulence by Changing the Plate-Liquid Contact Angles}

\author{Hao-Ran Liu\aff{1},
  Kai Leong Chong\aff{2},
  Chong Shen Ng\aff{1},
  Roberto Verzicco\aff{3,4,1}
 \and Detlef Lohse\aff{1,5,}\corresp{\email{d.lohse@utwente.nl}}}

\affiliation{\aff{1}Physics of Fluids Group and Max Planck Center for Complex Fluid Dynamics, MESA+ Institute and J. M. Burgers Centre for Fluid Dynamics, University of Twente, P.O. Box 217, 7500AE Enschede, The Netherlands
\aff{2}Shanghai Key Laboratory of Mechanics in Energy Engineering, Shanghai Institute of Applied Mathematics and Mechanics, School of Mechanics and Engineering Science, Shanghai University, Shanghai, 200072, PR China
\aff{3}Dipartimento di Ingegneria Industriale, University of Rome “Tor Vergata”, Via del Politecnico 1, Roma 00133, Italy
\aff{4}Gran Sasso Science Institute - Viale F. Crispi, 7 67100 L’Aquila, Italy
\aff{5}Max Planck Institute for Dynamics and Self-Organization, Am Fassberg 17, 37077 G\"ottingen, Germany}

\begin{document}

\maketitle

\begin{abstract}
This numerical study presents a simple but extremely effective way to considerably enhance heat transport in turbulent wall-bounded multiphase flows, 
namely by using oleophilic walls. As a model system, we pick the Rayleigh-B\'enard setup, filled with an oil-water mixture. For oleophilic walls, using only $10\%$ volume fraction of oil in water, we observe a remarkable heat transport enhancement of more than $100\%$ as compared to the pure water case. 
In contrast, for oleophobic walls, the enhancement is only of about $20\%$ as compared to pure water. The physical explanation of the heat transport increment for oleophilic walls is that thermal plumes detach from the oil-rich boundary layer and carry the heat with them: 
In the bulk, the oil-water interface prevents the plumes from mixing with the turbulent water bulk and to diffuse their heat. To confirm this physical picture, we show that the minimum amount of oil necessary to achieve the maximum heat transport is set by the volume fraction of the thermal plumes. Our findings provide guidelines of how to optimize heat transport in wall-bounded thermal turbulence. Moreover, the physical insight of how coherent structures are coupled with one of the phases of a two-phase system has very general applicability for controlling transport properties in other turbulent wall-bounded multiphase flows.

\end{abstract}


\section{Introduction}
The key property of a turbulent flow is its ability to efficiently transport heat, mass, and/or momentum. Understanding this property allows to control the global transport, enhance or reduce it, which is not only of fundamental interest, but also highly relevant in various industrial applications. For example one can think of heat transfer enhancement for cooling applications \citep{dhi98} or drag reduction in pipe or channel flow \citep{bushnell2015arfm,chung2021arfm}. A typical way to control integral properties is by manipulating coherent turbulent structures, which play a crucial role in the global transport \citep{smi11,holmes2012book,haller2015arfm,graham2021arfm}. An example is in thermally driven turbulence where the global heat transport can be enhanced by manipulating the coherent thermal plumes -- a major heat carrier in  thermal convection \citep{ahl09,loh10,chi12,shishkina2021prf}. 

Indeed, previous studies have proposed many different methods to enhance heat transport related to thermal plume manipulation: One method is to promote the detachment of plumes from the boundary layers by adding surface roughness \citep{she96,cil99,du00,ahl09,sal14,wag15,xie2017,zhu17b,jiang2018} or adding shear \citep{bergman2011book,pirozzoli2017,blass2020,wang2020-bofu}. Another method is to confine the system in span-wise direction  in order to increase the coherence of the thermal plumes and thus the heat transport \citep{huang2013prl,chong2015prl,cho17}.

However, also the addition of a new phase can  enhance heat transport. For example, the mechanical injection of air bubbles \citep{gvozdic2018jfm,ng2020jfm}  or nucleating vapor bubbles by boiling \citep{dhi98,zho09,bif12,lakkaraju2013,guzman2016} are both quite efficient heat transfer enhancement strategies. While in the former case this is purely due to the extra mechanical stirring by the rising buoyant bubbles, in the latter case the vapor bubbles also act as direct heat carriers. But as shown by Wang {\it et al.} \citep{wang2019nc-ziqi,wang2020ijhmt-ziqi}, a considerable heat transfer can also emerge when the second phase has comparable density, even for small volume fraction of the second phase. 

In this study, we present another, novel, and simple way to substantially enhance heat transfer in wall-bounded thermally driven two-phase turbulence, namely by manipulating the wettability of the walls. We then reveal the underlying physics of the enhancement and in particular the crucial role of the interaction between the coherent structures (plumes) and the second phase. As paradigmatic example to study turbulent heat transfer in wall-bounded systems, we pick Rayleigh-B\'enard (RB) convection \citep{ahl09,loh10,chi12,shishkina2021prf}. The system consists of a two-phase fluid in a box heated from below and cooled from above. Next to water, we use oil as a second phase, with its higher heat-conduction properties as compared to water. In our setup the top and the bottom walls are either both oleophilic or both oleophobic. With oleophilic walls, we reveal a novel flow dynamics, namely an ``expressway" of oil for the thermal plumes. When thermal plumes detach from the oil-rich thermal boundary layers, they can be transported together with the oil phase. These oil-rich thermal plumes act as the dominant heat transport phase, bypassing the water phase. Thus the global heat transfer is significantly enhanced. We further reveal a strong correlation between the oil phase and the thermal plumes in the oleophilic cases, and show that the minimum amount of oil to achieve maximum heat transport is set by the volume fraction of thermal plumes.

The organization of this paper is as follows. The numerical methodology is introduced in Section \ref{meth}. The main results on turbulent RB convection with multiphase is presented in Section \ref{dis}. The mechanism of heat transfer enhancement is revealed in Section \ref{mech}. The paper ends with conclusions and an outlook.

\begin{figure}
\centering
\includegraphics[width=1\linewidth]{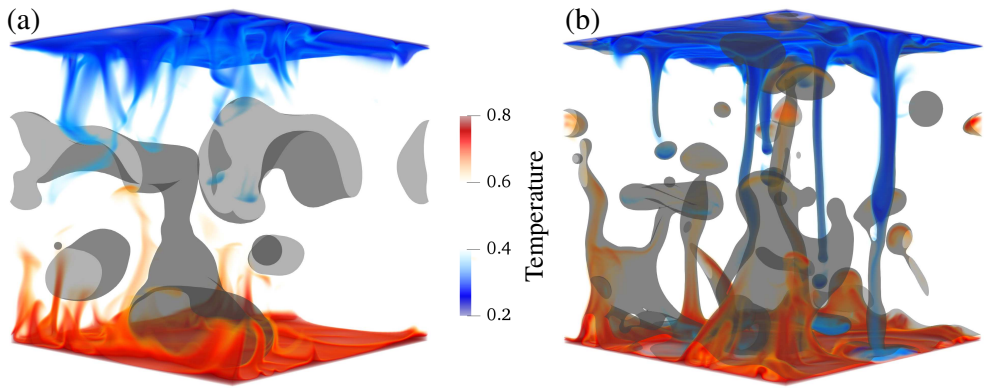}
\caption{\label{fig-3d}   The volume rendering of thermal plumes (red and blue) and oil-water interface (gray) with the (a) oleophobic and (b) oleophilic walls, respectively, for volume fraction of oil $\alpha=10\%$ at $\Ra_o=4\times10^8$ in oil and $\Ra_w=10^8$ in water. The non-dimensional heat transfer (the Nusselt number) is $\Nun=34.2$ in (a), $\Nun=50.3$ in (b), $\Nun_w=32.8$ in pure water, and $\Nun_o=49.2$ in pure oil. The corresponding movies are shown as Supplementary Material.}
\end{figure}

\section{Methodology}\label{meth}

The numerical method used here combines the phase-field model \citep{jaqcmin1999jcp, ding2007jcp, liu2015jcp,soligo2021jfe} and a direct numerical simulations solver for the Navier-Stokes equations namely \href{https://github.com/PhysicsofFluids/AFiD}{AFiD}, which is a second-order finite-difference open-source solver \citep{ver96, poe15cf}. This method is developed to simulate the turbulence with two immiscible fluids and well validated in \cite{liu2021jfm,liu2021jcp} by comparing the results to previous experimental, theoretical, and numerical results. The mass conservation is ensured by using the method of \cite{wang2015jcp}. Various test cases in multiphase turbulence are extensively discussed in \cite{liu2021jcp}.

\subsection{Governing equations}

In the phase-field model, the evolution of the volume fraction of water, $C$, is governed by the Cahn-Hilliard equation,
\begin{equation}
\frac {\partial C} {\partial t} + \nabla \cdot ({\bf u} C) = \frac{1}{\Pe}\nabla^2 \psi,
\label{ch}
\end{equation} 
where $\bf u$ is the flow velocity, and $\psi= C^{3} - 1.5 C^{2}+ 0.5 C  -\Cn^{2} \nabla^2 C$ the chemical potential. We set the P\'eclet number $\Pe=0.9/\Cn$ and the Cahn number $\Cn=0.75h/H$ according to the criteria proposed in \citep{ding2007jcp, liu2015jcp}, where $h$ is the mesh size and $H$ the domain height.

The flow is governed by the Navier-Stokes equations, the heat transfer equation and the incompressibility condition,
\begin{equation}
\rho \left(\frac {\partial {\bf u}} {\partial t} + {\bf u} \cdot \nabla {\bf u}\right)= - \nabla P + \sqrt {\frac{\Prn}{\Ra}} \nabla \cdot [\mu (\nabla {\bf u}+ \nabla {\bf u}^{T})] + {{\bf F}_{st}}+{\bf G},
\label{ns}
\end{equation}

\begin{equation}
\rho c_p \left(\frac{\partial {\theta}}{\partial t} + {\bf u} \cdot \nabla \theta \right) =   \sqrt{\frac{1}{ \Prn \Ra }} \nabla \cdot (k \nabla \theta),
\label{t}
\end{equation}

\begin{equation}
\nabla \cdot {\bf u}= 0,
\label{con}
\end{equation}
where $\theta$ is the dimensionless temperature. We define all dimensionless fluid properties (indicated by $q$), including the density $\rho$, the dynamic viscosity $\mu$, the kinematic viscosity $\nu=\mu/\rho$, the thermal conductivity $k$, the thermal expansion coefficient $\beta$, the thermal diffusivity $\kappa$, and the specific heat capacity $c_p=k/(\kappa\rho)$, in a uniform way as follows,
\begin{equation}
q=C+\lambda_q(1-C),
\end{equation}
where $\lambda_q=q_o/q_w$ is the ratio of the material properties of oil and water. Here the surface tension force is defined as ${\bf F}_{st}=6\sqrt{2}\psi \nabla C / (\Cn \We)$, and the gravity ${\bf G}=[C+\lambda_\beta \lambda_\rho (1-C)] \, \theta  {\bf j}$ with $\bf j$ being the unit vector in vertical direction. Besides the property ratios, the global dimensionless numbers controlling the flow are $\Ra= \beta_w g H^3 \Delta /(\nu_w \kappa_w)$, $\Prn=\nu_w / \kappa_w$, and $\We=\rho_w U^2 H/\sigma$, where $\Delta$ is the temperature difference between the bottom and top plates and $U=\sqrt{\beta_w g H \Delta}$ the free-fall velocity, and $\sigma$ the surface tension coefficient. The response parameter is the Nusselt number $\Nun=QH/(k_w\Delta)$ with $Q$ being the dimensional heat transfer. Note that the governing equations are normalized by the properties of water. Therefore, only global dimensionless numbers (e.g. $\Ra$ and $\Prn$, without subscripts) and the ratios $\lambda_q$ of the material properties show up in eqs.~(\ref{ns})~(\ref{con}). More numerical details can be found in \citep{liu2021jcp,liu2021jfm}.

\subsection{Configurations of turbulent Rayleigh-B\'enard convection with two phases}

In this study, the control parameters are the volume fraction of the oil, $\alpha$, the wettability of the wall surface (oleophobic and oleophilic), and the ratio of the Rayleigh number $\Ra_o/\Ra_w$. Here the subscripts ``o" and ``w" respectively indicate oil and water. The ratio $\Ra_o/\Ra_w$ is varied by changing the ratio of thermal expansion coefficients, $\lambda_\beta$, that of the thermal conductivity coefficients, $\lambda_k$, and that of the dynamic viscosity, $\lambda_\mu$. Here $\Ra_o>\Ra_w$ is used since the oil used has higher heat-conduction properties. Similarly to in the single-phase RB convection, we found that $\Nun$ always depends on the Rayleigh number, which depends on combinations of the fluid properties. Therefore, $\Ra_o/\Ra_w$ is used here to characterize the ratio of fluid properties of oil and water. We take the density ratio $\lambda_\rho$ as $1$ in an attempt to separate the various effects. Here we restrict the density variations only to temperature (i.e. oil and water are assumed to have the same density at the same temperature) and focus on the effects of $\beta$, $k$ and $\mu$. Note that the viscosity is assumed to be independent on temperature for simplicity (Oberbeck-Boussinesq approximation \citep{obe79,bou03}). 
Other parameters are kept fixed, including the Weber number $\We$ (the ratio of inertia to surface tension), $\Ra_w$ (the dimensionless temperature difference between the plates) and the Prandtl numbers $\Prn_w$ and $\Prn_o$ (a material property), at values allowing considerable turbulence and based on the properties of water and oil. 
We therefore fixed $\Ra_w=10^8$, $\Prn_o=\Prn_w=4.38$, and $\We=800$. This value is e.g. obtained by choosing the realistic values of $\rho_w=1000~kg/m^3$, $\sigma=0.021~N/m$, $U\approx0.4~m/s$, and $H\approx0.1~m$. We varied $1 \le \Ra_o/\Ra_w \le 12$. The latter value is motivated by $\Ra_o/\Ra_w=11.8$ as obtained in a system consisting of \href{https://www.shinetsusilicone-global.com/catalog/pdf/kf96_e.pdf}{KF-96L-1cs silicone oil} and water. Next we varied the volume fraction $0\% \le \alpha \le 100\%$ and the wettability (oleophilic or oleophobic), where we only took the two extreme contact angles $0^\circ$ and $180^\circ$.

The simulations were performed in a three-dimensional cubic domain of size $H^3$. Periodic boundary conditions were used in the horizontal directions, and no-slip walls at the top and bottom. The oleophobic and oleophilic conditions of the wall surface were realized by the Dirichlet boundary condition used in the phase-field model. 
We use a multiple resolution strategy \citep{liu2021jcp} in the simulation: the uniform mesh with $576^3$ gridpoints to capture the two phases and the stretched mesh with $288^3$ gridpoints for the velocity and temperature fields. The mesh is sufficiently fine and is comparable to corresponding single-phase flow studies \citep{ste10,poe13}.

\section{Heat transfer in the RB convection with oleophilic and oleophobic walls}\label{dis}

\begin{figure}
\centering
\includegraphics[width=1\linewidth]{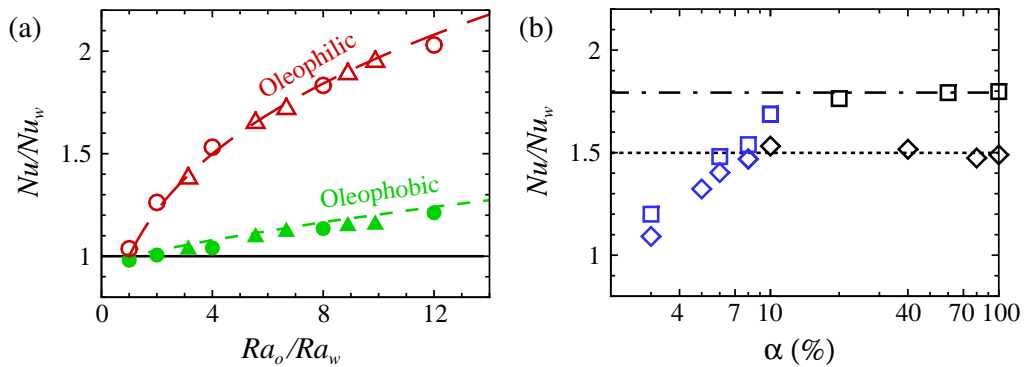}
\caption{\label{fig-nu}  Heat transfer in terms of $\Nun/\Nun_w$ as function of (a) $\Ra_o/\Ra_w$ for $\alpha=10\%$, and (b) of $\alpha$ at $\Ra_o/\Ra_w=4$ and $\Ra_w=10^8$ ($\diamondsuit$), and $\Ra_o/\Ra_w=8$ and $\Ra_w=10^7$ ($\square$). Empty symbols denote the convection with the oleophilic walls, and filled ones with the oleophobic walls. In (a), $\Ra_o$ is varied by changing $\beta_o$ ($\bigcirc$) and by changing $\mu_o$ and $k_o$ ($\triangle$), and $\Ra_w$ is kept constant at $=10^8$. The three lines represent the predictions by the Grossmann-Lohse theory \citep{ste13} with $\Ra_o$ of $100\%$ oil (red long dashed line), $\Ra_w$ of $100\%$ water (black solid line) and $\Ra_{eff}=\alpha \Ra_o+(1-\alpha) \Ra_w$ (green dashed line). In (b), the black symbols denote the cases with $\Nun$ in agreement with the prediction of the Grossmann-Lohse theory with $100\%$ oil (dash-dotted line corresponding to data with $\square$ and dotted line to $\diamondsuit$), and the blue ones with $\Nun$ more than $2\%$ smaller than the prediction.}
\end{figure}

Initially, one big oil drop is placed at the center of the domain full of water. Then the oil drop breaks up into many smaller ones due to the thermally driven turbulent convection. As shown in figure \ref{fig-3d}(a), the oil phase is repelled by the oleophobic walls, whereas in figure \ref{fig-3d}(b), the oil phase is attracted by the oleophilic boundaries. This preferential wetting causes the boundary layer region to be occupied by water in the former and by oil in the latter case. These different near-wall flow structures significantly affect the heat transfer through the flow. In order to further elucidate this phenomenon, we have performed simulations with oleophobic and oleophilic walls, respectively, at various ratio $\Ra_o/\Ra_w$ from $1$ to $12$ for fixed $\Ra_w=10^8$ and $\alpha=10\%$, as shown in figure \ref{fig-nu}(a). Since we define $\Ra_o>\Ra_w$, for all cases, both with the oleophobic walls and with the oleophilic walls, the heat transport is enhanced as compared to the case with pure water, but the amount of enhancements for the oleophilic cases is significantly larger than for the oleophobic cases.

In the oleophobic case, the amount of heat transfer enhancement linearly depends on $\Ra_o/\Ra_w$, and an enhancement of $21\%$ is achieved at $\Ra_o/\Ra_w=12$. Based on the observation, we define an effective Rayleigh number as $\Ra_{eff} = \alpha \Ra_o+(1-\alpha) \Ra_w$. With this definition, the values of $\Nun$ for the oleophobic cases well agree with the Grossmann-Lohse (GL) \citep{gro00,gro01,ste13} prediction for $Nu(Ra_{eff})$. 

Next, we examine the more intriguing case with oleophilic walls. Surprisingly, the heat transfer enhancement in the oleophilic cases dramatically increases up to $103\%$ at $\Ra_o/\Ra_w=12$ (see figure \ref{fig-nu}a), which is $4$ times larger than the enhancement in the oleophobic cases for the same parameters. It is remarkable to observe that the heat transfer efficiency can be as large as the cases with $100\%$ oil, despite only $10\%$ volume fraction of oil has been used.

One question naturally arises: what is the minimum amount of oil required for the system to transfer the maximum heat, assumed as the value for $100\%$ oil? To answer this question, we performed simulations with oleophilic walls for various $\alpha$ from $0\%$ to $100\%$ at fixed $\Ra_o/\Ra_w=4$ and $\Ra_w=10^8$. For $\alpha<8\%$, $\Nun$ in the oleophilic cases increases with $\alpha$, while for $\alpha$ larger than this critical value ($\alpha_c \approx 8\%$), the heat transfer increase saturates to a value close to $\Nun_o$ (the value for $100\%$ oil) within $\pm 2\%$, as shown in figure \ref{fig-nu}(b). The same result is also achieved at $\Ra_o/\Ra_w=8$ and $\Ra_w=10^7$, but then with $\alpha_c \approx 10\%$.

\begin{figure}
\centering
\includegraphics[width=1\linewidth]{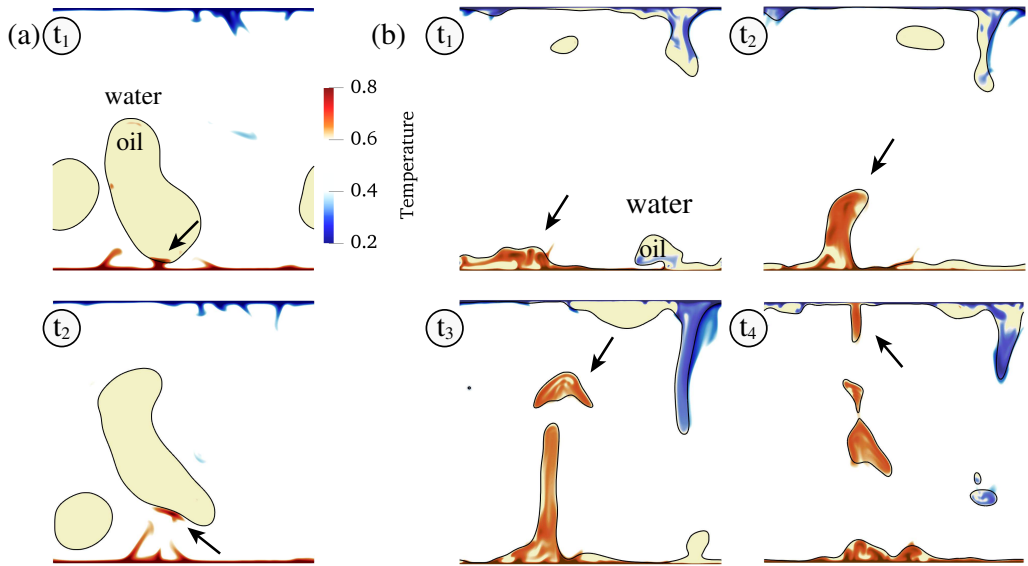}
\caption{\label{fig-2d}   Visualization of temperature fields and the oil-water interface for $\alpha=10\%$ at $\Ra_o/\Ra_w=4$ and $\Ra_w=10^8$ with the (a) oleophobic and (b) oleophilic walls. 
The color legend is red/blue for hot/cold plumes. The black lines represent the oil-water interface and oil is in color yellow. With oleophobic walls (a), most thermal plumes travel in the water phase, whereas with oleophilic walls (b), they are carried by the oil phase. For the shown times of the snapshots it holds $t_{n+1}>t_n$.}
\end{figure}

\section{Mechanism of heat transfer enhancement}\label{mech}

What causes the significant enhancement of heat transfer in the cases with oleophilic walls despite using only a small amount of oil? The major reason is related to the thermal plumes, which are the primary heat carrier in thermal turbulence \citep{ahl09,loh10,chi12,shishkina2021prf}. To show the effect of the thermal plumes, we visualize them in figure \ref{fig-2d}(a) for the oleophobic cases, and in figure \ref{fig-2d}(b) for the oleophilic cases. 
{\color{black}In the cases with oleophobic walls,} the thermal plumes mainly stay outside the oil phase after travelling into the bulk. The reason is that surface tension prevents the entrainment of external fluid into the oil phase. The only way for the heat carried by plumes to be transported across the interface of the two immiscible fluids is via thermal diffusion, the effects of which however are sufficiently small at such high $\Ra$. 

In the oleophobic cases the thermal plumes are mainly exposed to the turbulent bulk. In contrast, in the oleophilic cases, the thermal plumes detach from the oil-rich boundary layer (figure \ref{fig-2d}b). At the same time, the oil-water interface not only keeps thermal plumes inside the oil phase, but also prevents the heat carried by plumes to be mixed into the turbulent bulk. As a result, most hot (cold) thermal plumes can travel to the opposite plate with little heat loss (gain) to the turbulent bulk as they are thermally strongly coupled to the oil phase. As shown in figure \ref{fig-2d}(b), the oil carrying the hot plumes rises up in the shape of column, then breaks up into drops, and eventually coalesces with the oil in the upper boundary layer. During this process, the oil builds up an ``expressway" to transfer the heat efficiently. Thus, the heat transfer in the oleophilic cases is always significantly enhanced up to the same value as in the system with $100\%$ oil, and only a small amount of oil (larger than $\alpha_c$) is required due to the strong coupling of the oil and the thermal plumes.

\begin{figure}
\centering
\includegraphics[width=1\linewidth]{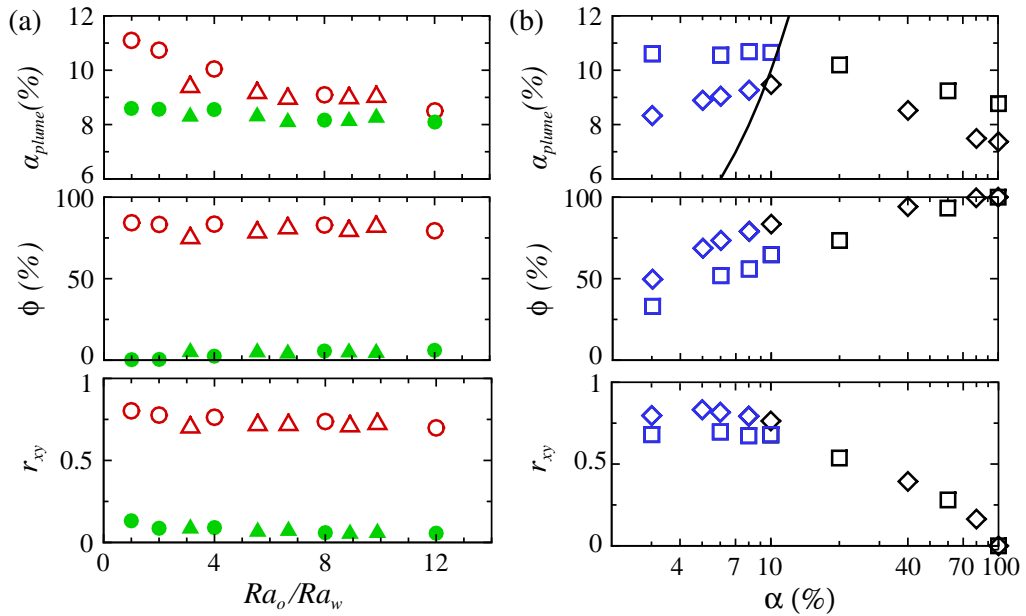}
\caption{\label{fig-data}   Volume fraction of thermal plumes, $\alpha_{plume}$, the ratio of thermal plumes in oil over the total thermal plumes, $\phi$, and the correlation coefficient, $r_{xy}$ with $x$ being the thermal plumes and $y$ the oil phase, (a) as function of $\Ra_o/\Ra_w$ for $\alpha=10\%$, and (b) as function of $\alpha$ at $(\Ra_w,\Ra_o)=(10^7,8\times 10^7)$ and $(\Ra_w,\Ra_o)=(10^8,4 \times 10^8)$. Symbols denote the same cases as in figure \ref{fig-nu}, and the line in (b) represents $\alpha_c=\alpha_{plume}$ (Eq.~\ref{ac}).}
\end{figure}

To quantitatively confirm this physical picture, we measure the amount of thermal plumes, using as thermal plumes definition condition $|T- \langle T \rangle|>\sqrt{\langle(T-\langle T\rangle)^2\rangle}$ with $T({\bf x},t)$ being the local temperature and $\langle \rangle$ the spatial and temporal average. 
In figure \ref{fig-data}, we show the volume fraction of the total thermal plumes in the domain, $\alpha_{plume}$, the ratio of thermal plumes in oil over the total thermal plumes, $\phi$, and the correlation coefficient $r_{xy}=\sum(x-\bar{x})(y-\bar{y})/\sqrt{\sum(x-\bar{x})^2\sum(y-\bar{y})^2}$ between the distribution of thermal plumes ($x$) and oil ($y$) in the whole domain, where the bar represents the spatial average and $r_{xy}$ ranges from $0$ (no correlation) to $1$ (perfect correlation).

For the oleophilic cases, for various $\Ra_o/\Ra_w$ and fixed $\alpha=10\%$ (figure \ref{fig-data}a), we observe $\phi=80\%\pm5\%$ and $r_{xy}=0.75\pm0.05$, whereas for the oleophobic cases we always have $\phi<6\%$ and $r_{xy}<0.09$. This finding shows that, in the oleophilic cases, most thermal plumes indeed are strongly coupled to the oil, contrasting to the weak coupling in the oleophobic cases. 


Finally how to determine the critical volume fraction of oil $\alpha_c$ to achieve maximum heat transfer efficiency? In figure \ref{fig-data}(b), we show the oleophilic cases at various $\alpha$ and fixed $(\Ra_w,\Ra_o)=(10^7,8\times 10^7)$ and $(10^8,4 \times 10^8)$, and the critical condition for cases with $\Nun$ smaller than or close to $\Nun_o$, 
\begin{equation}
\alpha_c=\alpha_{plume},
\label{ac}
\end{equation}
which well agrees with our numerical results. For $\alpha<\alpha_{plume}$ (blue symbols in figure \ref{fig-data}b), although oil and the thermal plumes are still well coupled ($r_{xy}\approx0.75$), there is not enough oil to generate all the plumes ($\phi\approx50\%$), leading to $\Nun$ smaller than $\Nun_o$ (figure \ref{fig-nu}b). On the other hand, for $\alpha>\alpha_{plume}$ (black symbols in figure \ref{fig-data}b), the excess amount of oil does not contribute to generating additional thermal plumes, and thus the heat transfer enhancement levels off and $\Nun$ remains close to $\Nun_o$ as shown in figure \ref{fig-nu}(b). Therefore, $\alpha_c$ is clearly determined by the volume of thermal plumes in the domain. 


\section{Conclusions and outlook}\label{conc}
We have shown a novel way to significantly enhance the heat flux in wall-bounded thermally driven turbulence, namely by adding an oil phase (even of relatively small volume fraction) to the water and at the same time using oleophilic walls. The heat transport enhancement is brought about by the newly-found flow structure, where the thermal plumes are strongly coupled with the oil phase for cases with oleophilic walls. With the thermal plumes fully encapsulated inside the oil phase, the maximum enhancement can be obtained. 

Our method and findings can easily and directly be applied and extended to other wall-bounded turbulent multiphase
transport systems relevant in  science and technology, way beyond thermal transport, namely to the turbulent multiphase mass or 
momentum transfer, such as drag in pipe or channel flow of oil-water mixtures \citep{pal1993aiche,roccon2021jfm}, or of bubbly flow \citep{lu05,san06,cec10,elbing2013,murai2014}, and angular momentum transfer in turbulent multiphase Taylor-Couette flow \citep{ber07,mur08,gil13,spandan2018jfm,sri15,hu2017sa,yi2021jfm,bakhuis2021prl}. The reason is that what we manipulate by the wall coating and the qualitative and quantitative choice of the two phases are the coherent structures of the turbulent multiphase flow, and these determine not only the heat transfer in turbulent flow, but also mass and momentum transfer and therefore the overall drag.

\section*{Supplementary movies}
Supplementary  movies  are  available at \href{https:}{URL}

\section*{Acknowledgments}
The work was financially supported by ERC-Advanced Grant under the project no.~$740479$. We acknowledge PRACE for awarding us access to MareNostrum in Spain at the Barcelona Computing Center (BSC) under the project $2020225335$ and $2020235589$. This work was also carried out on the national e-infrastructure of SURFsara, a subsidiary of SURF cooperation, the collaborative ICT organization for Dutch education and research.

\section*{Declaration of interests}
The  authors  report  no  conflict  of  interest.


\end{document}